# Slow Spin Relaxation in Single Endohedral Fullerene Molecules

Jie Li, Lei Gu and Ruqian Wu*

*Department of Physics and Astronomy, University of California, Irvine, California 92697-4575, USA.*

**ABSTRACT:** Well-protected magnetization, tunable quantum states and long coherence time are desired for the use of magnetic molecules in spintronics and quantum information technologies. In this work, endohedral fullerene molecules M@$C_{28}$ with different transition metal cores were explored through systematic first-principles calculations and spin dynamics analyses. Many of them have stable magnetization, giant magnetic anisotropy energy and bias-tunable structure. In particular, some of them may have coherence time up to several milliseconds for their quantum spin states at high temperature (~10K) after full consideration of spin-vibration couplings. These results suggest that these M@$C_{28}$ provide a rich pool for the development of single-molecule magnets, magnetic molecular junctions, and molecular qubits.

* E-mail: wur@uci.edu.

# I. INTRODUCTION

The foremost important issue for using magnetic molecules as units in spin filtering, quantum computing and magnetic sensing devices is to have a long quantum coherence time [1-3]. For example, well characterized quantum states and long coherence times are among the critical prerequisites proposed by DiVincenzo for qubits [4]. It is recognized that spins in a magnetic molecule may provide quantum states with energy splittings in the microwave range for quantum computation and hence research interest in magnetic molecules has noticeably surged in recent years [5-9]. Different strategies have been proposed for the design of magnetic molecules with large spin relaxation time ($\tau$), from single core magnetic molecules [9-10], to multiple core magnetic clusters [11-13]. In general, spin relaxation time has an activated Arrhenius-like behavior at a high temperature: i.e., $\tau = \tau_0 e^{U_{eff}/k_B T}$, where $\tau_0$ represents the inverse attempt frequency, and $U_{eff}$ and $k_B$ are the effective barrier for relaxation and the Boltzmann constant, respectively [14]. Without the vibration-assisted tunneling effect, $U_{eff}$ is solely determined by the magnetic anisotropy energy (MAE). This inspired an active search for molecules with enhanced MAEs, using low-symmetry structures [15-18] and heavy elements including rare earth and actinide atoms [19-21]. Nevertheless, recent studies indicated that spin-vibration coupling, hyperfine interaction, and intermolecular interaction may even more significantly affect the quantum coherence of magnetic molecules [22], particularly when their MAEs are high. It is timely critical to establish roles of different factors in the molecular quantum decoherence and, as a further step, to find molecules with long quantum coherence time through the manipulation of magnetization, spin orbit coupling and spin-vibration coupling (SVC).

In this work, we systematically investigate the spin relaxation in a series of bias-controllable magnetic endohedral fullerenes: M@$C_{28}$ (M=3d, 4d and 5d transition metals), through a combination of density functional calculations and spin dynamics simulations. Several such endohedral fullerenes (e.g., Ti@$C_{28}$, Zr@$C_{28}$, and U@$C_{28}$) have already been synthesized in recent experiments [23] and hence there should be no obvious technological barrier for developing similar molecules if attractive properties are predicted. The closed mesh topology of fullerene shells offers a protective environment for magnetic atoms and reduce spin-vibration coupling, especially for the low energy modes. We find that these endohedral fullerenes have various attractive properties such as structural bistability, ferroelectricity, multiple magnetic phases with large MAEs. In particular, some of them have long coherence time for their low-energy Kramers doublets, up to the millisecond order even at a temperature of ~10 K. These findings suggest a possibility of designing single magnetic molecules for quantum computing, sensing and spintronic applications.

## II. METHODOLOGY

All ab initio calculations in this work were carried out with the Vienna ab-initio simulation package (VASP) at the level of the spin-polarized generalized-gradient approximation (GGA) [24]. The interaction between valence electrons and ionic cores was considered within the framework of the projector augmented wave (PAW) method [25,26]. The energy cutoff for the plane wave basis expansion was set to 500 eV. A Hubbard $U_{eff}$ = U-J = 2.0 eV within the Dudarev scheme was added to take account of the onsite coulomb interaction for transition metals atom d orbitals [27]. All atoms were fully relaxed using the conjugated gradient method for the energy minimization until the force on each atom became smaller than 0.01 eV/Å, and $10^{-6}$ as the convergence criteria for total energy was selected for all DFT calculations. The density functional perturbation theory is carried out within the density functional framework to calculate the vibrational properties [28]. The spin-vibration dynamics and the spin-spin dynamics calculations were simulated by using the master equations with DFT parameters, (see details in the Supplemental Material [29]).

### III. RESULTS AND DISCUSSION

$C_{28}$ has a tetrahedral structure as Fig. 1, with three pentagons directly fused on the dome [30]. Experimental investigations indicated that $C_{28}$ can be stabilized by insertion of appropriate core atom [23]. The ab initio molecular dynamics (AIMD) simulations furtherly show their dynamic stability. In the AIMD simulations, gap phase Ir@$C_{28}$ and Rh@$C_{28}$ are not noticeably deformed after 10ps at 300K as shown in the videos in supplemental Material [29]. Our density functional theory (DFT) calculations indicate that most gas-phase M@$C_{28}$ molecules (M= 3d, 4d and 5d transition metals) have two structural phases (denoted as I and II below), both with a $C_3$ symmetry as shown in Fig. 1. The displacements of M atoms between the two phases are shown in Table 1. Using Ir@$C_{28}$ as an example, Ir atom displaces by 1.1 Å as the phase changes and the $C_{28}$ cage also deforms slightly, i.e., the aspect ratios (height vs. width) change from 0.860 in the pristine $C_{28}$ to 0.923 (phase I) and 0.996 (phase II), respectively. The planar average charge density difference in Fig. 1 shows that electrons transfer from Iridium to carbon atoms. The Bader charge of Iridium atom is -1.1e (phase I) or -0.9e (phase II), and charge redistribution is detailed in Fig. S3 [29]. The separation of positive and negative charge centers gives rise to dipole moments of 0.326eÅ and -0.701eÅ in these two phases. In Table I, one may see that dipole moments of all M@$C_{28}$ are nonzero. This offers a possibility for gate control between their phases and possibly strong response to microwave manipulation especially in the THz regime.

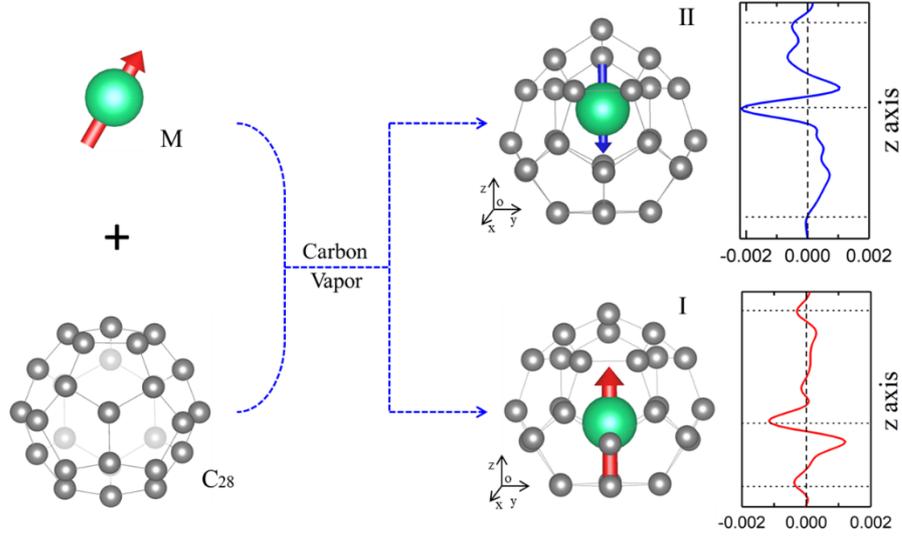

FIG. 1. (Left) Schematic diagram of the synthesis process of M@$C_{28}$ in two phases (I and II) (Red and blue arrows represent the direction of the dipoles). (Right) The corresponding charge difference of Ir@$C_{28}$, i.e., in I and II phases, respectively. (Positive and negative represent charge accumulation and depletion, respectively).

It is desired to have large MAEs, as the first step, to search magnetic molecules with long coherence time. Systematic DFT calculations show that most of M@$C_{28}$ are magnetic except for the Ti, Zr and Hf cases as shown in Fig. 2a. Taking Ir@$C_{28}$ as the example, one may see a strong hybridization between Ir and $C_{28}$ orbitals from the curves of projected density of states (PDOS) of Ir@$C_{28}$ in Fig. 2b. Significant spin-polarization is also induced around C atoms, especially for those next to Ir as shown by the spin density in insets of Fig. 2b. Quantitatively, the total magnetic moments of Ir@$C_{28}$ are 5.0 $\mu_B$ and 1.0 $\mu_B$ in phases I and II, for which C atoms contribute 1.82$\mu_B$ and 0.46$\mu_B$, respectively. The relatively nonlocal magnetization in the carbon cage may induce weak spin polarization in supporting substrates such as graphene and hence offers a possibility to establish quantum entanglement among magnetic molecules. For example, a gate may control the density of carriers or energy alignment in supporting materials through which the long-range magnetic coupling between molecules can be either established or eliminated.

*Table I. The displacements (Δu) of M atoms between phase transition and the difference of electric dipole moments ($\Delta D = D_I - D_{II}$, where $D_I$ and $D_{II}$ represent their electric dipole moments in phase I and II, respectively) of M@$C_{28}$.*

|          | Ti   | V    | Cr   | Mn   | Fe   | Co   | Ni   |
|----------|------|------|------|------|------|------|------|
| Δu (Å)   | 0.93 | 1.04 | 1.06 | 0.95 | 0.63 | 1.00 | 0.81 |
| ΔD (eÅ)  | 0.92 | 0.78 | 0.91 | 0.66 | 0.40 | 0.61 | 0.44 |
|          | Zr   | Nb   | Mo   | Tc   | Ru   | Rh   | Pd   |
| Δu (Å)   | 0.02 | 0.20 | 0.83 | 1.08 | 1.02 | 0.47 | 0.39 |
| ΔD (eÅ)  | 0.01 | 0.46 | 0.80 | 1.46 | 1.34 | 0.48 | 0.32 |
|          | Hf   | Ta   | W    | Re   | Os   | Ir   | Pt   |
| Δu (Å)   | 0.01 | 0.72 | 1.03 | 1.13 | 0.98 | 1.10 | 0.38 |
| ΔD (eÅ)  | 0.01 | 0.61 | 0.64 | 0.95 | 1.37 | 1.02 | 0.33 |

To determine their MAEs, we calculate the torque $T(\theta)$ as a function of the polar angle $\theta$ between the magnetization and the z-axis. The torque is defined as [31,32],

$$T(\theta) = \frac{dE(\theta)}{d\theta} = \sum_{occ} \left\langle \psi_{jk} \left| \frac{\partial H_{so}}{\partial \theta} \right| \psi_{jk} \right\rangle \quad (1)$$

with the spin-orbit coupling Hamiltonian $H_{so} = \sum_i \xi_i \hat{l}_i \cdot \hat{s}_i$ with $\hat{l}_i = -ir_i \times \nabla$. The indices $i$ and $j$ refer to individual atoms and eigenstates, and the summation in Eq. 1 goes over all occupied states. By integrating $T(\theta)$, we may obtain the total energy $E(\theta)$ as a function of the polar angle $\theta$, and the MAEs equal to the energy differences between the lowest and highest energies. In Fig. 2c, Ir@$C_{28}$ has the lowest and highest energies occurring at $\theta=0°$ and $\theta=90°$, indicating that its easy axis is along the z-axis for both phase I and phase II. Nevertheless, the energy differences, $MAE = E(\theta = 90°) - E(\theta = 0°)$, drastically differ, from 32.2 meV for the phase I to only 0.93meV for the phase II. To assess the tunability of this important parameter, we further calculate the Fermi level dependences of total and spin channel decomposed MAEs according to the rigid band model [31,33] and results are shown in Fig. 2d. For both cases, the MAE curves are flat in a broad energy range around the Fermi level, but the MAE of the phase I may drop to a large negative value as the Fermi level moves down by 0.2 eV. This is because that the Fermi level skips through the hybridized $d_{z^2}$ orbital of Ir. In Fig. 2a, one may see that the 5d (4d) cores lead to larger MAEs than 4d (3d) cores by an order of magnitude. As we need molecules with large positive MAEs for applications, we choose Rh@$C_{28}$ and Ir@$C_{28}$ in phase I as examples for spin dynamic studies.

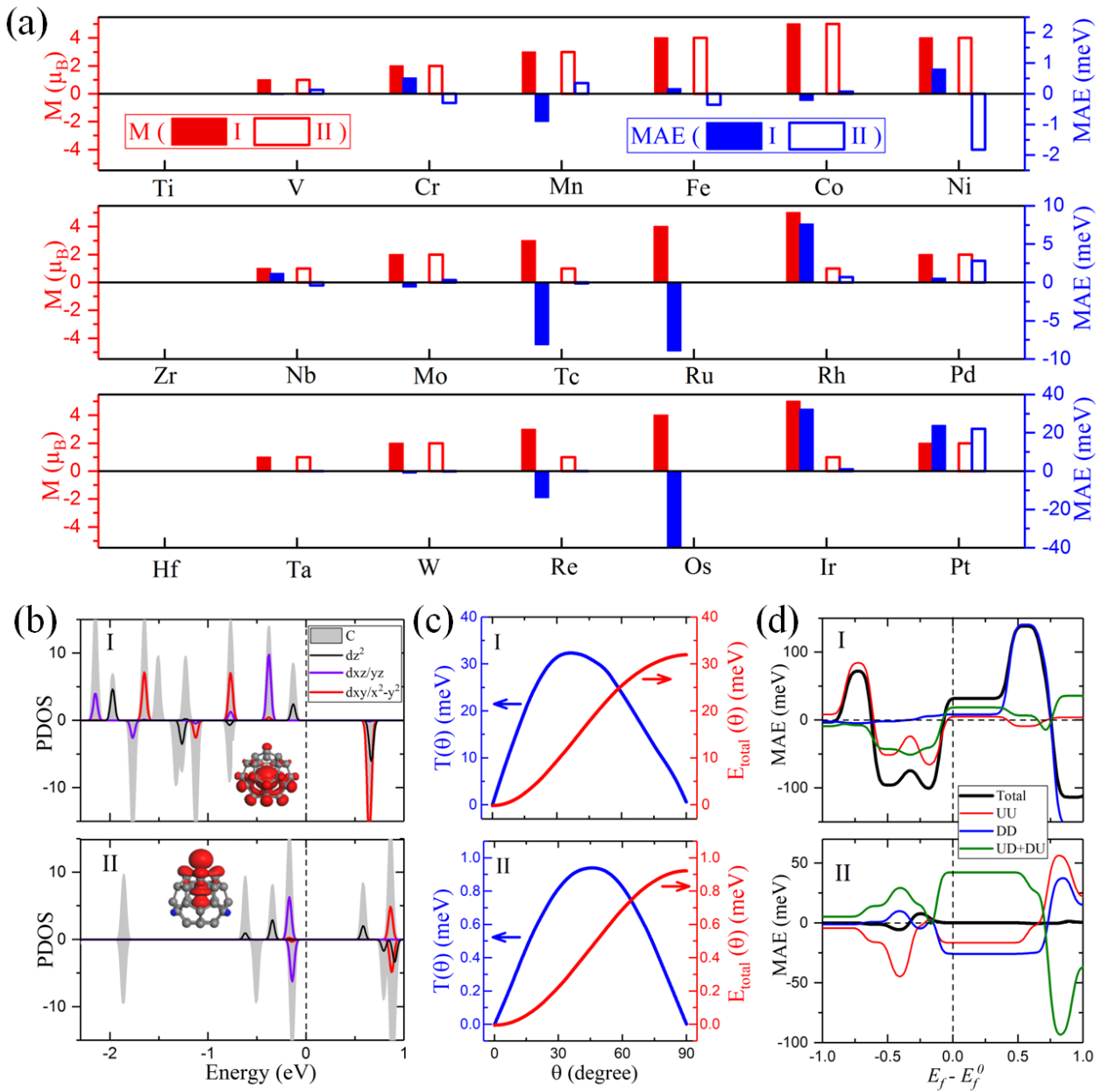

FIG. 2. (a) The calculated magnetic moments and MAEs of M@$C_{28}$ (M= 3d, 4d and 5d transition metals). (b) The PDOS of Ir@$C_{28}$ in two phases (inset are the corresponding spin density). (c)

*Calculated torque vs the angle θ for Ir@C$_{28}$. (d) The Fermi level dependent total and spin channel decomposed MAEs of Ir@C$_{28}$ from the rigid band model analyses.*

To describe quantum states of these magnetic molecule, we introduce a spin Hamiltonian as

$$H_{spin} = -D_{xx}S_x^2 - D_{yy}S_y^2 - D_{zz}S_z^2 - D_{xy}(S_xS_y + S_yS_x)$$

$$-D_{xz}(S_xS_z + S_zS_x) - D_{yz}(S_yS_z + S_zS_y) \quad (2)$$

where the magnetic anisotropy is extended to a D-tensor. These parameters are computed by the four-state mapping method [34]:

$$D_{\alpha\beta} = \frac{-1}{4S^2}(E_{\alpha\beta}^1 + E_{\alpha\beta}^4 - E_{\alpha\beta}^2 - E_{\alpha\beta}^3) \quad (3)$$

In which $E_{\alpha\beta}^{1-4}$ are the energies of four different spin configurations (see details in the Supplemental Material [29]). For a magnetic molecule with large $D_{zz}$, the energy diagram is sketched in Fig. 3(a). The lowest Kramers doublets can be used for quantum information operations. Since the MAE is high and so is the separation between different doublets, we may concentrate on the decoherence of the ground state doublet due to the SVC.

According to Eq. 2, the SVC is caused by vibration-induced changes of D parameters. Here, we investigate the effects of both the first- and second-order SVC coefficients on the spin decoherence. The first-order SVC coefficients can be obtained by taking the derivatives of Eq. 2 with respect to $u_{i\gamma}$ (the displacements of atom $i$ along $\gamma$ direction):

$$\frac{\partial H_{spin}}{\partial u_{i\gamma}} = \sum_{\alpha\beta}\frac{-\partial D_{\alpha\beta}}{\partial u_{i\gamma}}S_{\alpha\beta}^2 \quad (4)$$

with

$$\frac{\partial D_{\alpha\beta}}{\partial u_{i\gamma}} = \frac{-1}{4S^2}\left(\frac{\partial E_{\alpha\beta}^1}{\partial u_{i\gamma}} + \frac{\partial E_{\alpha\beta}^4}{\partial u_{i\gamma}} - \frac{\partial E_{\alpha\beta}^2}{\partial u_{i\gamma}} - \frac{\partial E_{\alpha\beta}^3}{\partial u_{i\gamma}}\right) \quad (5)$$

where, $-\frac{\partial E_{\alpha\beta}^n}{\partial u_{i\gamma}}$ ($n$=1,…,4) denotes the force acting on the atom $i$ along the $\gamma$ direction. The second-order SVC coefficients can be expressed as:

$$\frac{\partial^2 H_{spin}}{\partial u_{i\gamma}\partial u_{i'\gamma'}} = \sum_{ij}\frac{-\partial^2 D_{\alpha\beta}}{\partial u_{i\gamma}\partial u_{i'\gamma'}}S^2_{\alpha\beta} \qquad (6)$$

with

$$\frac{\partial^2 D_{\alpha\beta}}{\partial u_{i\gamma}\partial u_{i'\gamma'}} = \frac{-1}{4S^2}\left(\frac{\partial^2 E^1_{\alpha\beta}}{\partial u_{i\gamma}\partial u_{i'\gamma'}} + \frac{\partial^2 E^4_{\alpha\beta}}{\partial u_{i\gamma}\partial u_{i'\gamma'}} - \frac{\partial^2 E^2_{\alpha\beta}}{\partial u_{i\gamma}\partial u_{i'\gamma'}} - \frac{\partial^2 E^3_{\alpha\beta}}{\partial u_{i\gamma}\partial u_{i'\gamma'}}\right) \qquad (7)$$

In the DFT schemes with the plane wave bases, $-\frac{\partial E^n_{\alpha\beta}}{\partial u_{i\gamma}}$ and $\frac{\partial^2 E^n_{\alpha\beta}}{\partial u_{i\gamma}\partial u_{i'\gamma'}}$ (*n*=1,…,4) are connected to the Hellmann-Feynman forces and the Hessian matrices, respectively. The calculated vibration modes and energies of Rh@$C_{28}$ and Ir@$C_{28}$ in gas phase in the 0-50 meV range as shown in Fig. 3b. One can see that two lowest vibrations modes are in-plane, while the third one is along the z axis.

The dynamics of the quantum spin states can then be described by the master equation:

$$\frac{dp_m(t)}{dt} = \sum_n(p_n(t)q_{nm} - p_m(t)q_{mn}) \qquad (8)$$

where $p_m$ represent the probability of being at quantum spin states $|m\rangle$ and $q_{mn}$ represent the transition rate from quantum spin states $|m\rangle$ to $|n\rangle$. The magnetic relaxation pathways for a ground state doublet are those illustrated schematically in Fig. 3a, where both the Orbach and Raman processes are included. By using the master equations with DFT parameters [35-41], (see details in the Supplemental Material [29]), the probabilities of holding in the Kramers doublets of Ir@$C_{28}$ through relaxation are shown in Fig. 3c. While the excited doublet quickly decays (within $10^{-4}$ ms as shown in the inset), the ground state doublet decay rather slowly. One may see from Fig. 3c that a long spin-lattice relaxation time $T_1$ (2.3ms) can be achieved at a reasonably high temperature (10K) for the ground state doublet of Ir@$C_{28}$. For Rh@$C_{28}$ with weaker SOC, the SVC is weaker as well and the spin-lattice relaxation time $T_1$ is 0.7ms. To further explore the relevant decoherence processes of this single-molecule, the dephasing time $T_2$ of Ir@$C_{28}$ is also calculated as shown in the inset of Fig. 3c (see calculation details in the Supplemental Material [29]). One may see that $T_2$ of Ir@C$_{28}$ is long as well, ~1.6ms at a temperature of 10K. With both factors involved, the coherence time of the ground state Kramers doublet of Ir@C$_{28}$ is still at the level of 1 ms, exceeding the time scale for coherent manipulations of the electron spin (~10ns based on the existing apparatus) [42]. Obviously, the ground state doublet of Ir@C$_{28}$ has long coherence time after full consideration of spin-lattice relaxation and spin-spin dephasing. We would hope that our work may inspire experimental interest to synthesize Ir@C$_{28}$ or Rh@C$_{28}$ as they are promising for applications.

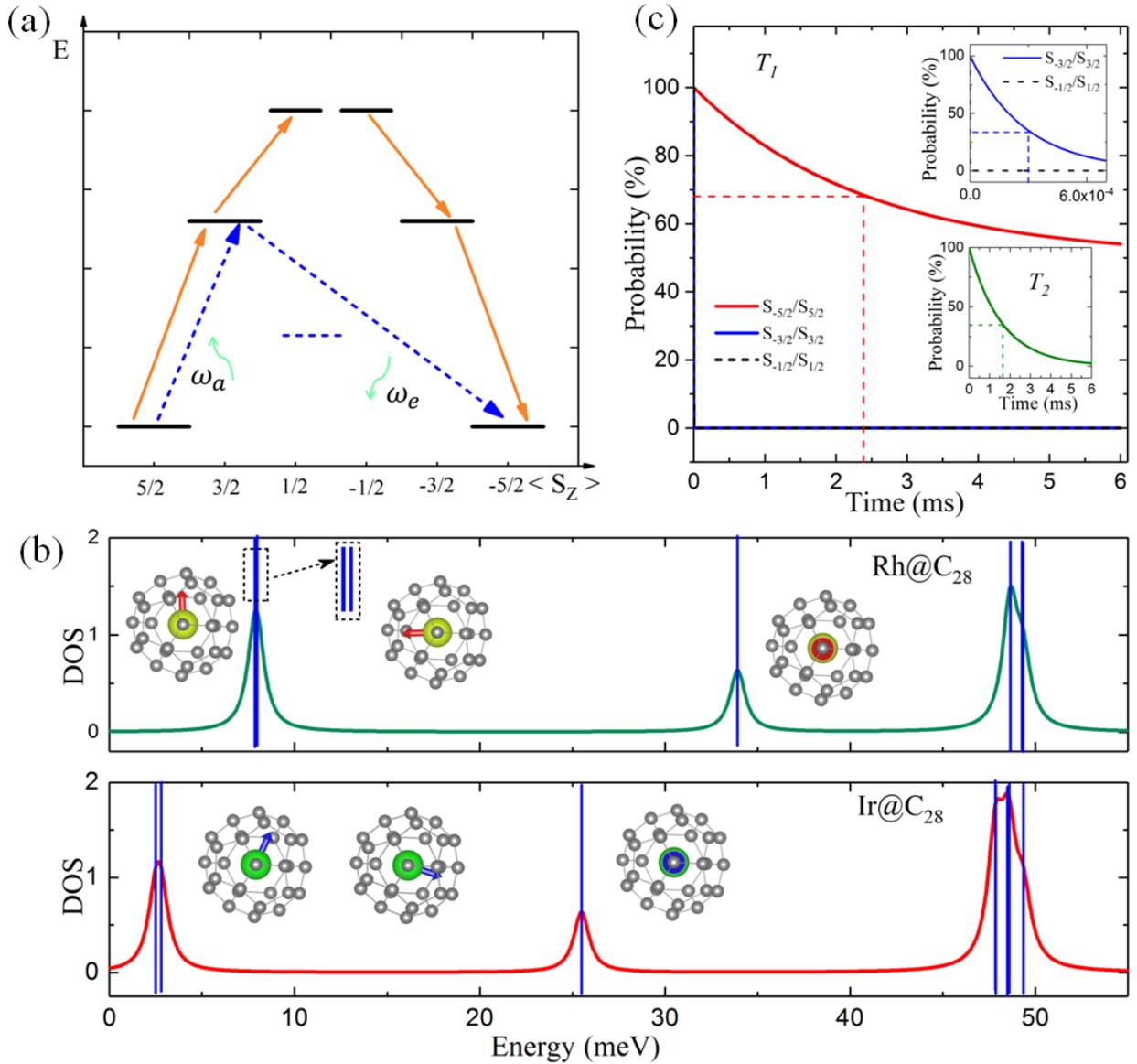

FIG. 3. (a) Schematics of the magnetic relaxation processes of spin states. The solid orange and dash blue arrows represent the Orbach process and Raman process; $\omega_a$ and $\omega_e$ represent the vibrational absorption and emission, respectively. The horizontal blue dashed line is a virtual state. (b) The phonon DOS of Rh@C$_{28}$ and Ir@C$_{28}$ with energies smaller than 50 meV from broadening of local vibrational modes (The blue lines are their corresponding vibration spectrum in gas phase). (c) The decaying of quantum spin states of Ir@C$_{28}$ in phase I (S=5/2) with time at a reasonably high temperature (10K). The bottom right inset is the decaying of the ground state doublet with time in T$_2$ mechanism (the phase coherence relaxation). The top right inset is the corresponding zoom-in decaying of the S$_{-3/2}$/S$_{3/2}$ and S$_{-1/2}$/S$_{1/2}$ in T$_1$ mechanism.

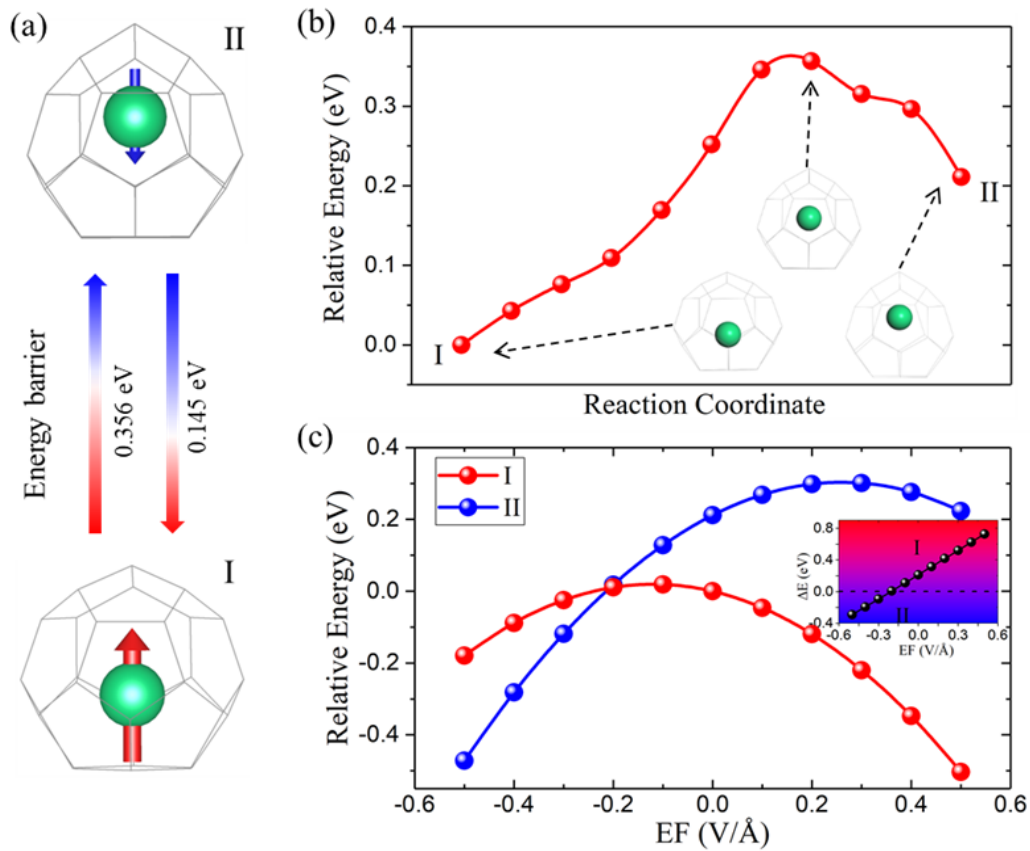

FIG. 4. (a) Schematic diagram of the switching between the phases I and II of Ir@C$_{28}$ (Red and blue arrows are represent the direction of the dipoles). (b) Energetics pathway for structural transition and the optimized structures for Ir@C$_{28}$ in phases I and II and the transition state between them. (c) The relative energy of Ir@C$_{28}$ in phases I and II as functions of the external electric field (negative external electric field is along z axis).

The other advantage of M@C$_{28}$ is that it has two stable phases with distinctly different quantum states as shown in Fig. 4a. From climbing image nudged elastic band (CINEB) calculations with 9 intermediate images between phases I and II, one may see that phase I (II) of Ir@C$_{28}$ has an energy barrier of ~ 0.356eV (0.145eV) in the switching process (see in Fig.4b). As Ir is ionized in the carbon cage, an external electric field may drive the phase transition. From the calculated total energies under an external electric field in Fig. 4c, one may convert Ir@C$_{28}$ from phase I to phase II by applying a reasonable electric field of 0.2V/Å. We may then tune the potential molecules in an array into different phases, as illustrated for two in Fig. S4 [29]. Graphene or silicon can be adopted as the substrate and Au electrodes can be buried underneath for gate bias.

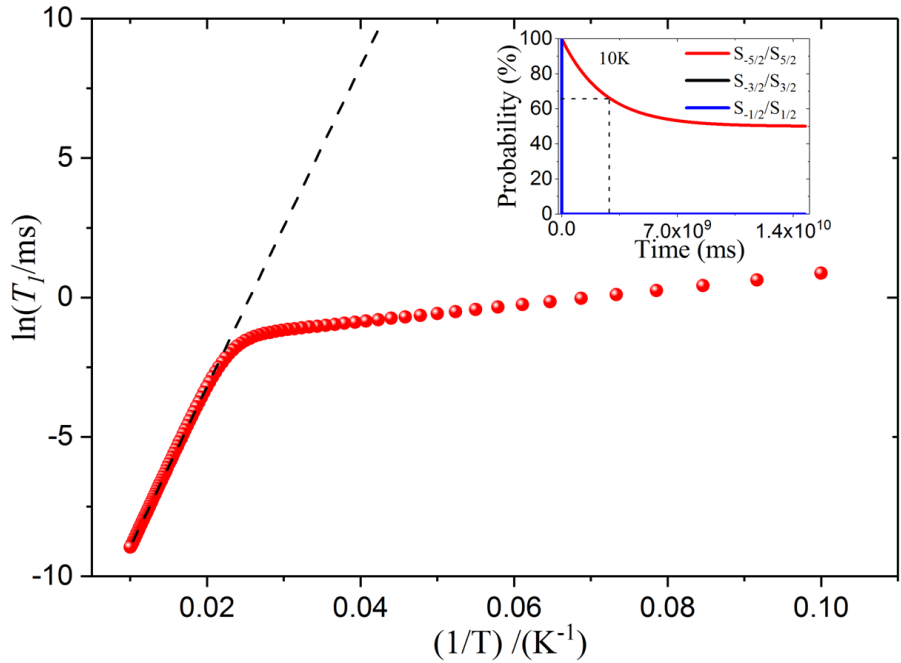

FIG. 5. The spin relaxation time as functions of temperature (the black dashed line is the Orbach barrier). The top right inset is the decaying of the ground state doublet of Ir@C$_{28}$ with only the Orbach processes be considered in $T_1$ mechanism (the spin-lattice relaxation) at 10K.

Note that Rh@@C$_{28}$ has smaller MAE and weaker SVC but has a similar relaxation time as

Ir@$C_{28}$. This shows that large MAE and weak SVC are the two equally important factors for the spin relaxation time of single magnetic molecules. Therefore, another strategy for the design of single-molecule magnets and qubits is to reduce the spin-vibration coupling. Meanwhile, our calculations show that the Orbach processes is much less important compared to the Raman processes for quantum decoherence at low temperature. For example, $T_1$ of the ground state doublet of Ir@$C_{28}$ in the inset of Fig. 5 can be as long as ~$2.6 \times 10^9$ ms if only the Orbach processes are considered at 10K. As the temperature increasing, the role of the Orbach processes will be furtherly enhanced, and the relaxation time then will have an Arrhenius behavior. Therefore, the Raman processes are dominant in the determination of coherence time of the ground state doublet a reasonably low temperature, especially for the magnetic molecules with large MAEs. Compared to other endohedral fullerenes, e.g., paramagnetic N@$C_{60}$ and P@$C_{60}$ molecules [43-45], the large MAE of Ir or Rh encapsulated $C_{28}$ may have reasonably high blocking temperature in the absence of high external magnetic field. While this work lays a conceptual groundwork, much more need to be done for the actual use of endohedral fullerene molecules in quantum information technologies, of course. Some other detrimental factors such as hyperfine interaction, intermolecular interaction and substrate effect should be examined.

## IV. CONCLUSION

In summary, we proposed that several endohedral fullerene molecule M@$C_{28}$ can have long quantum coherence time with full consideration of SVC, and hence can be developed for diverse applications such as qubit, spin filtering and quantum sensing. The closed mesh topology of $C_{28}$ offers a protective and low-symmetry environment for the magnetic atom, yet adequate exchange interaction among molecules for electric manipulation and the establishment of quantum entanglement. They also have sizeable electric dipole moments and dual stability, a leeway for bias control in quantum operations.

## Acknowledgements

Authors thank Profs. W. Ho and W. Evans at the University of California Irvine for insightful discussions. This work was supported by US DOE, Basic Energy Science (Grant No. DE-SC0019448). Calculations were performed on parallel computers at NERSC.

Modern Introduction (Cambridge University Press, New York, **2013**).